\documentclass[twocolumn,amssymb,nobibnotes,showpacs,superscriptaddress,aps,prl,longbibliography,floatfix]{revtex4-2}
\usepackage{amsmath,graphicx}
\usepackage[colorlinks,linkcolor=blue,urlcolor=blue,anchorcolor=blue,citecolor=blue]{hyperref}

\begin{document}


\title{Parametric interaction induced avoided dressed state crossings in cavity QED: generation of quantum coherence and equally weighted superposition of Fock states}


%
\author{L. L. Ping}
\affiliation{MOE Key Laboratory of Advanced Micro-Structured Materials,
	School of Physics Science and Engineering, Tongji University, Shanghai, China 200092}
\author{W. Li}
\affiliation{MOE Key Laboratory of Advanced Micro-Structured Materials,
	School of Physics Science and Engineering, Tongji University, Shanghai, China 200092}
\author{C. J. Zhu}
\email[Corresponding author:]{cjzhu@tongji.edu.cn}
\affiliation{School of Physical Science and Technology, Soochow University, Suzhou 215006, China}

\author{Y. P. Yang}
\email[Corresponding author:]{yang\_yaping@tongji.edu.cn}
\affiliation{MOE Key Laboratory of Advanced Micro-Structured Materials,
	School of Physics Science and Engineering, Tongji University, Shanghai, China 200092}
\author{G. S. Agarwal}
\email[Corresponding author:]{girish.agarwal@tamu.edu}
\affiliation{Institute for Quantum Science and Engineering, and Departments of Biological and Agricultural Engineering and Physics and Astronomy, Texas A\&M University, College Station, Texas 77843, USA}


\date{\today}

\begin{abstract}
We present a new paradigm in the field of cavity QED by bringing out remarkable features associated with the avoided crossing of the dressed state levels of the Jaynes Cummings model. We demonstrate how the parametric couplings, realized by a second order nonlinearity in the cavity,  can turn the crossing of dressed states into avoided crossings.  We show how one can generate coherence between the avoided crossing of dressed states. Such coherences result, for example, in quantum beats in the excitation probability of the qubit. The quality of quantum beats can be considerably improved by adiabatically turning on the parametric interaction. We show how these avoided crossings can be used to generate superpositions of even or odd Fock states with the remarkable property of equal weights for the states in superposition. The fidelity of generation is more than 95\%. In addition, we show strong entanglement between the cavity field and the qubit with the concurrence parameter exceeding 90\%.
\end{abstract}

\pacs{}

\maketitle

\section{Introduction}
Avoided level crossing~\cite{heiss1990avoided,eleuch2013avoided} is the phenomenon where two energy levels can't be pushed through each other and has been an important branch of physics. This phenomenon was firstly discovered in 1932 by studying a two level system in quantum mechanics~\cite{zener1932non,landau1932theorie}. Suppose that there are two energy levels labeled by $E_1$ and $E_2$, respectively. In the absence of external perturbation these two levels would have crossed if the original energy states were degenerate, i.e., $\Delta E=E_1-E_2=0$. However, in the presence of a perturbation on a two level system, the energy exchange between two states take place. Therefore, the eigenvalues of the system will not become degenerate but will have a hyperbolic shape where the minimal energetic distance is proportional to the perturbation strength~\cite{rotter2001dynamics}. The phenomenon of level repulsion is called as the avoided level crossing, which has been widely used in many different quantum systems, including atoms~\cite{rubbmark1981dynamical,rapol2002precise}, semiconductor devices~\cite{patterson1988muonium} and other systems~\cite{berns2008amplitude,wiersig2006formation,heiss1990avoided,davis1986random,hornbostel1989application,eleuch2013avoided}, which can be described as a ``general" two-state system with some couplings. In cavity QED systems, the avoided level crossing can be achieved by using the atom-cavity coupling strength. The avoided level crossing then becomes the famous vacuum Rabi splitting~\cite{haroche2020cavity,agarwal2012quantum,sanchez1983theory,agarwal1984vacuum,boca2004observation,maunz2005normal,yoshie2004vacuum}.

In the avoided level crossing region, the two states become strongly mixed, yielding various interesting phenomena. For example, a direct and natural consequence of the avoided crossing is the entanglement behavior~\cite{bell2002generation,karthik2007entanglement,pachniak2021creation,de2021entanglement}. This has been exploited to some extent in earlier works that seek to create entangled states such as the W or GHZ state by using the superpositions of two states that develop at avoided crossings~\cite{bruss2005multipartite,unanyan2002entanglement,braungardt2007error}. Moreover, the Landau-Zener tunneling between two energy levels takes place if these two levels of a time-dependent Hamiltonian are avoided crossing~\cite{rubbmark1981dynamical,agarwal1994realization,damski2006adiabatic,huang2011landau}. A Berry phase is accumulated in addition to a dynamical phase if an eigenstate encircles adiabatically degeneracy points~\cite{oh2008entanglement}. The first order quantum phase transition, an abrupt change in the ground state of a many body system as parameters of a system vary, is related with the avoided crossings of two lowest energy levels~\cite{damski2006adiabatic}.

In this paper, we present a new paradigm in the field of cavity QED by bringing out new effects associated with the avoided crossing of the dressed state levels of the Jaynes Cummings (JC) model. We demonstrate how the parametric couplings can turn the crossing of dressed states into avoided crossings. Note that this is different from the avoided crossings mentioned earlier in the context of the vacuum Rabi splittings. We show how one can generate coherence between the avoided crossing of dressed states. Such coherences can be monitored via the quantum beats in the excitation probability of the qubit. These avoided crossings can be used to generate superpositions of even or odd Fock states with the remarkable property of equal weights for the states in superposition. Besides, there is strong entanglement between the cavity field and the qubit. The qubit could be an atom or superconducting qubit or a quantum dot. As an example we can produce superpositions of photonic Fock states like $|0\rangle+i|2\rangle$, $|0\rangle+i|4\rangle$ and $|1\rangle+i|3\rangle$~\cite{note2}.

\begin{figure}[ht!]
	\centering
	\includegraphics[width=\linewidth]{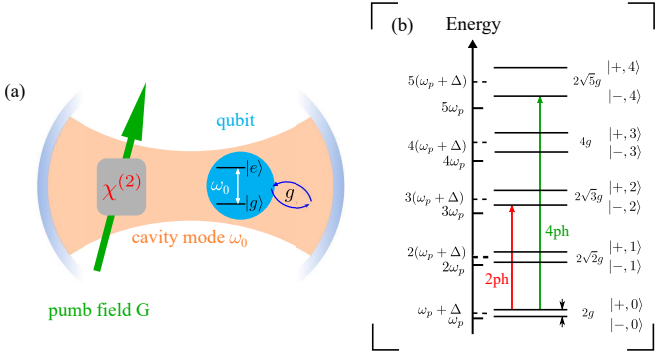}
	\caption{(a) Schematic diagram of a single qubit cavity QED system where the cavity photons are generated via the second-order nonlinearity with nonlinear interaction strength $G$ induced by a strong pump field. Here, the qubit transition frequency $\omega_0$ is identical to the cavity resonant frequency, and the qubit-cavity coupling strength is denoted by $g$. (b) Dressed state picture of the system where $|\pm,n\rangle\equiv(|e,n\rangle\pm|g,n+1\rangle)/\sqrt{2}$. The red and green arrows correspond to the transitions involving two- and four-photon processes, respectively.}~\label{Figs:Fig1}
\end{figure}
To begin with, we consider the standard JC model where a single two-level qubit with transition frequency $\omega_0$ is trapped in a single mode cavity with resonant frequency $\omega_0$ identical to qubit frequency. In addition as shown in Fig.~\ref{Figs:Fig1}(a), the qubit-cavity coupling strength is denoted by $g$, and the cavity is driven by pumping at frequency $\omega_P$, a  second-order nonlinear crystal with the nonlinear interaction strength $G$, corresponding to the well known optical parametric amplification process. In a frame rotating with frequency $\omega_P/2$, the Hamiltonian of the system can be written as
\begin{equation}~\label{Hsys}
	H=\Delta(a^\dag a + \sigma_{+}\sigma_{-})+g (a\sigma_{+}+a^\dag\sigma_{-})+G(a^2+a^{\dag 2})
\end{equation}
where $a^\dag(a)$ is the creation (annihilation) operator of the cavity mode, and $\sigma_+=|e\rangle\langle g|$ and $\sigma_-=\sigma_+^\dag$ are the spin raising and lowering operators of the qubit. Here, $\Delta=\omega_0-\omega_P/2$ is the detunings of the cavity and the qubit with respect to $\omega_P/2$. The structure of the spectrum of the eigenstates of Eq.~(\ref{Hsys}) is discussed in Refs.~\cite{tomka2014exceptional,gutierrez2021probing}. For many other important outcomes of Eq.~(\ref{Hsys}) in different contexts see Refs.~\cite{zhu2020squeezed,huang2009enhancement,huang2009normal,agarwal2016strong,leroux2018enhancing,qin2018exponentially,agarwal1990cooperative,qin2021generating,burd2021quantum}. Obviously, in the absence of the driving field, i.e., $G = 0$, the system goes back to a typical JC model. However, in the presence of the optical parametric amplification, i.e., $G\neq0$, all the new physical effects arise which we discuss in detail below. Before we discuss full numerical results, we like to highlight the physics behind the dressed state crossing and the formation of avoided crossings.

For $G=0$, as shown in Fig.~\ref{Figs:Fig1}(b), the eigenstates are the well known doublets in JC model, i.e., $|\pm,n\rangle\equiv(|e,n\rangle\pm|g,n+1\rangle)/\sqrt{2}$ with eigenvalue $\lambda_{\pm,n} \equiv (n+1)\Delta\pm\sqrt{n+1}g$. Obviously, the energy of eigenstate $|+,n\rangle$ increases, but the energy of eigenstate $|-,n\rangle$ decreases as the qubit-cavity coupling strength $g$ is enhanced. Thus, two eigenstates will cross at a specific detuning which is locked with the coupling strength. In the presence of the perturbation, i.e., the pump field, avoided level crossing between dressed states will occur, yielding quantum beats, entanglement and the superposition state of even or odd Fock states. We present a simple, rather approximate discussion to bring out how the avoided crossings result.

Consider the dressed state $|+,0\rangle$ and $|-,2\rangle$ with energies $\lambda_{+,0}=\Delta+g$, $\lambda_{-,2}=3\Delta-\sqrt{3}g$. These two dressed states cross at $\lambda_{+,0}=\lambda_{-,2}$ for $\Delta=(1+\sqrt{3})g/2$. Let us now assume that $G$ is small so that we can retain the coupling of $|+,0\rangle$ and $|-,2\rangle$ as illustrated by the red arrow in Fig.~\ref{Figs:Fig1}(b). Note that $|+,0\rangle=(|e,0\rangle+|g,1\rangle)/\sqrt{2}$ and $|-,2\rangle=(|e,2\rangle-|g,3\rangle)/\sqrt{2}$ are coupled in two different ways via the parametric drive which causes two photon transitions $|e,0\rangle\overset{\sqrt{2}G}{\Longleftrightarrow}|e,2\rangle$ and $|g,1\rangle\overset{\sqrt{6}G}{\Longleftrightarrow}|g,3\rangle$. This coupling results in two new energy levels which are now separated at $\Delta=(1+\sqrt{3})g/2$ by $2|\langle +,0|G(a^2+a^{\dag 2})|-,2\rangle|=\sqrt{2}(\sqrt{3}-1)G$. This example demonstrates in a simple way how the parametric drive can produce avoided crossing of the dressed states $|+,0\rangle$ and $|-,2\rangle$. 
\begin{figure}[ht!]
	\centering
	\includegraphics[width=\linewidth]{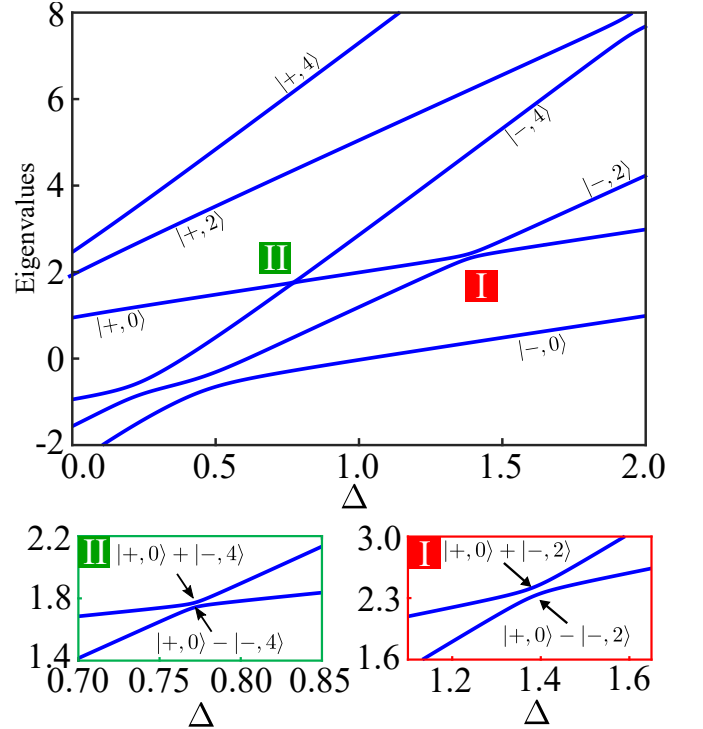}
	\caption{Eigenvalues of the system against the detuning $\Delta$ with system parameters $g=1$ and $G=0.1$. Avoided level crossing involving $|+,0\rangle$ and $|-,2\rangle$ ($|+,0\rangle$ and $|-,4\rangle$) is indicated by label I (II). The lower two plots are the zoom-ins near the avoided level crossing regimes.}~\label{Figs:Fig1b}
\end{figure}
In Fig.~\ref{Figs:Fig1b} we show the behavior of the eigenvalues of the Hamiltonian, i.e., Eq.~(\ref{Hsys}), as a function of $\Delta$ for $g=1$, $G=0.1$. We have kept the parametric coupling law so that we could work with truncated Hilbert space up to $5$ photons. we display two avoided crossings involving dressed states $|+,0\rangle$ and $|-,2\rangle$ ($|+,0\rangle$ and $|-,4\rangle$). 

\begin{figure}[ht!]
	\centering
	\includegraphics[width=\linewidth]{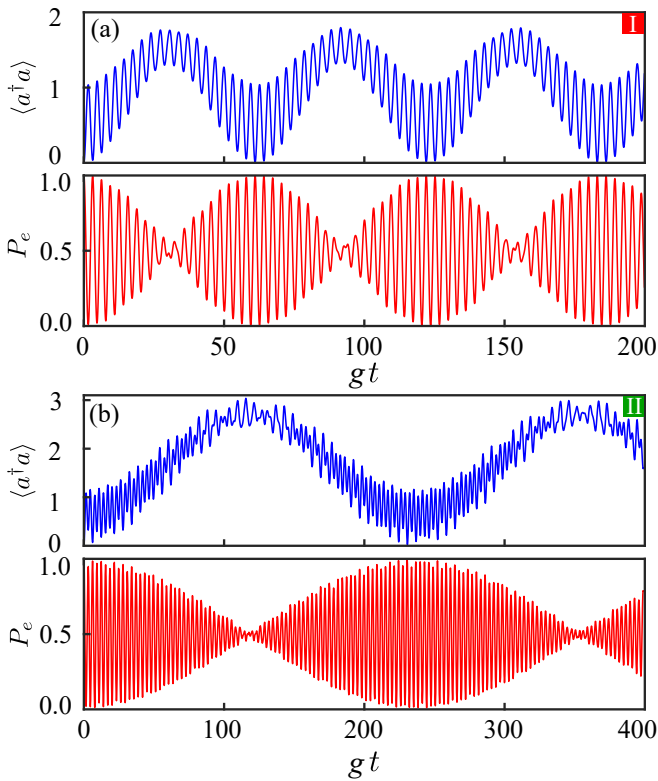}
	\caption{Dynamical evolution of the mean photon number $\langle a^\dag a\rangle$ (blue) and probability of finding qubit in the excited state $P_e$ (red) with initial condition $\Psi(0)=|e,0\rangle$. Here, we choose the detuning $\Delta=(1+\sqrt{3})g/2$ and other system parameters are given by $g=1.0$ and $G=0.1$.}~\label{Figs:Fig2}
\end{figure}
To quantify the above physical discussion, we numerically solve the Schr\"{o}dinger equation $\partial\Psi/\partial t=-iH\Psi$ with $\Psi$ being the wave function of the system. In Fig.~\ref{Figs:Fig2}, we plot the mean photon number $\langle a^\dag a\rangle$ and the probability of finding a qubit in the excited state $P_e$ as a function of the normalized evolution time $g t$. Here, we choose the detuning $\Delta=(1+\sqrt{3})g/2$ (i.e., the avoided level crossing I) and other system parameters are the same as those used in Fig.~\ref{Figs:Fig1b}. As shown in Fig.~\ref{Figs:Fig2}, there exist two different oscillation frequencies in time evolution of $\langle a^\dag a\rangle$ and $P_e$. The faster oscillation results from the effective qubit-cavity coupling with a short period of $T_1=\pi/g$, while the slower oscillation originates from the energy difference at the avoided level crossing with a long period of $T_2=2\pi/[(\sqrt{6}-\sqrt{2})G]$ (the beat frequency is defined as $f_{\rm beat}=1/T_2$). At the quiet spots, the mean photon number reaches its maximum, but the population of the qubit excited state $P_e=0.5$. Likewise, at the avoided level crossing II (i.e., $\Delta=(1+\sqrt{5})g/4$), quantum beat behavior can also be observed in the cavity excitation spectrum and population of the qubit excited state (see the supplementary material for details). 
\begin{figure}[ht!]
	\centering
	\includegraphics[width=\linewidth]{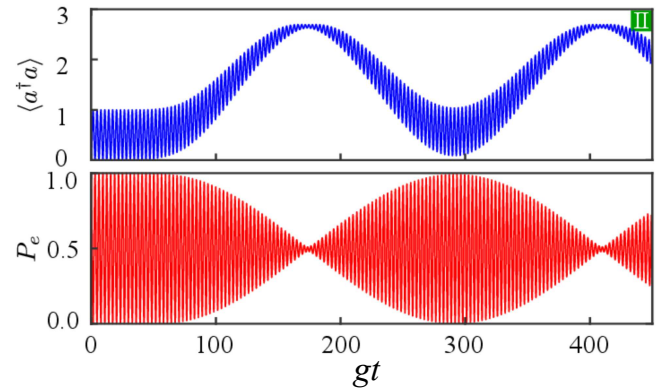}
	\caption{The density matrix of the intracavity photon state (a) and the Wigner function (b) by numerically solving the Schr\"{o}dinger equation. Here, the evolution time $g t\approx31$ and the detuning is chosen as $\Delta=(1+\sqrt{3})g/2$, corresponding to the avoided level crossing I.}~\label{Figs:Fig3}
\end{figure}
Next, let's study the properties of photon states and the entanglement between qubit and cavity photons at the quiet spot of quantum beats, where the coherence of two dressed states are maximally established. Thus, the system is evolved into the superposition state of two dressed states, i.e., $\Psi\approx(|e\rangle|{\rm even}\rangle+{\rm e}^{i\phi}|g\rangle|{\rm odd}\rangle)/{\cal A}$ with $\phi$ being the relative phase and ${\cal A}$ being the normalized coefficient. Here, the states $|{\rm even}\rangle$ and $|{\rm odd}\rangle$ denote the superposition of Fock states with even or odd photon numbers, respectively. For the crossing point I, the even and odd states can be approximately expressed as $|{\rm even}\rangle=(|0\rangle+i|2\rangle)/\sqrt{2}$ and $|{\rm odd}\rangle=(|1\rangle+i|3\rangle)/\sqrt{2}$ according to the numerical solutions of the Schr\"{o}dinger equation. In Fig.~\ref{Figs:Fig3}(a), we show the density matrix of the cavity photon state with system parameters $g=1.0$ and $t\approx31/g$, where the superposition of states $|+,0\rangle$ and $|-,2\rangle$ are perfectly generated by the coherence. Here, the initial condition is chosen as $|e,0\rangle$ and the detuning $\Delta$ is chosen near the avoided level crossing point I. As shown in Fig.~\ref{Figs:Fig3}(a), the probabilities of finding the Fock states $|0\rangle$, $|1\rangle$, $|2\rangle$ and $|3\rangle$ are approximately equal to each other. At the same time, the  coherences between Fock states $|0\rangle$ and $|2\rangle$ ($|1\rangle$ and $|3\rangle$) denoted by diagonal bars in panel (a) reach their maximum. In Fig.~\ref{Figs:Fig3}(b), we show the corresponding Wigner function of the cavity photons, which exhibits a similar pattern of cat state, where the negativities (one in the origin and other two in the second and fourth quadrants) of the Wigner function represent the nonclassicality of the cavity photon state and the quantum interference fringes are denoted as a signature of coherence of Fock states. In the following, we will discuss how even or odd photon number Fock states result in the negativities in the Wigner function. It should be noted that the even and odd states that we generate are different from the CAT states involving squeezed coherent states which are generated in the ultrastrong coupling regime~\cite{gutierrez2021probing,leroux2018enhancing,qin2018exponentially}.


\begin{figure}[ht!]
	\centering
	\includegraphics[width=\linewidth]{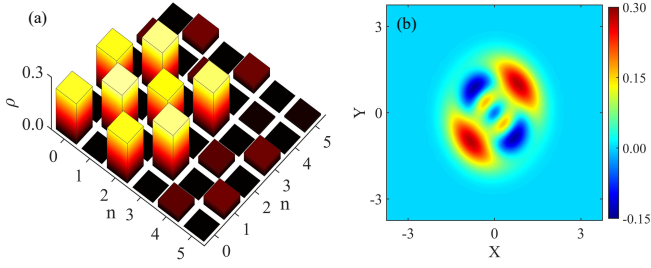}
	\caption{(a,d) Photon state distributions, (b,e) the Wigner functions by solving the Schr\"{o}dinger equation and (c,f) the analytical Wigner functions. Panels (a-c) correspond to the case where the qubit excited state $|e\rangle$ is detected. Panels (d-f) show the case of detecting the qubit ground state. Here, the evolution time $g t\approx31$ and the detuning is $\Delta=(1+\sqrt{3})g/2$, corresponding to the avoided level crossing I.}~\label{Figs:Fig3n}
\end{figure}
With the coherence between dressed states, the qubit and the cavity photons are evolved into the maximally entangled state simultaneously. As shown in Fig.~\ref{Figs:Fig3n}(a), if the qubit excited state $|e\rangle$ is detected, only Fock states with even photon numbers can be observed with equal weights. In this case, the Wigner function shown in panel (b) exhibits a similar pattern of even cat state, where two negativities occur in the second and fourth quadrants symmetrically. To show the physical mechanism more clearly, we analytical calculate the Wigner function $W(\alpha)=(1/\pi^2)\int d^2\lambda C_w(\lambda)\exp{(\alpha\lambda^*-\alpha^*\lambda)}$ by assuming the photon state $|{\rm ph}\rangle=|{\rm even}\rangle$. Here, the characteristic function $C_w(\lambda)={\rm Tr}[\rho \hat{D}(\lambda)]$ with density matrix $\rho=|{\rm ph}\rangle\langle {\rm ph}|$ and the displacement operator $\hat{D}(\lambda)=\exp{(\lambda a^\dag-\lambda^* a)}$. Integrating the characteristic function $C_w$ by assuming $\lambda=x'+iy'$, one can obtain
\begin{equation}
	W(\alpha)=\frac{2}{\pi}{\rm e}^{-2|\alpha|^2}[(1-2|\alpha|^2)^2+4\sqrt{2}{\rm Re}(\alpha){\rm Im}(\alpha)].
\end{equation}
Note that the first term in the square bracket is alway positive and the value of the Wigner function at the origin is always positive, the negativity of the Wigner function appears in the regime of Re$(\alpha){\rm Im}(\alpha)<-(1-2|\alpha|^2)^2/(4\sqrt{2})<0$. As shown in Fig.~\ref{Figs:Fig3n}(c), two negativities of this analytical Wigner function also appear in the second and fourth quadrants which matches well with the numerical result. On the contrary, the cavity field will collapse to the superposition of odd photon number Fock states with equal weights [see Fig.~\ref{Figs:Fig3n}(d)] if the qubit is measured in its ground state. The corresponding Wigner function is shown in panel (e) by solving the Schr\"{o}dinger equation, which is similar to the odd cat state. Likewise, the Wigner function can also be calculated analytically by assuming the density matrix $\rho=|{\rm odd}\rangle\langle {\rm odd}|$, which yields
\begin{eqnarray}
	W(\alpha)&=&\frac{1}{\pi}e^{-2|\alpha|^2}\left[-2(1-4|\alpha|^2)^2 +8|\alpha|^4(1+\frac{4}{3}|\alpha|^2)\right.\nonumber\\
	& & \left.+8\sqrt{6}(\frac{4}{3}|\alpha|^2-1){\rm Re}(\alpha){\rm Im}(\alpha)\right].
\end{eqnarray}
Obviously, the analytical Wigner function shown in panel (f) agrees well with the numerical one. It is clear to see that the Wigner function at the origin is negative. We note that interference fringes in the Wigner functions [see panels (b) and (e)] prove the coherence between even or odd photon number Fock states, while the negativity of the Wigner function indicates the nonclassciality of superposition states of even or odd photon number Fock states.

\begin{figure}[ht!]
	\centering
	\includegraphics[width=\linewidth]{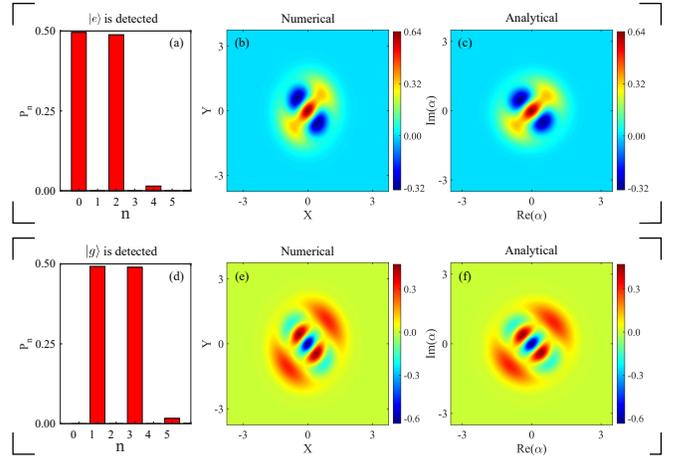}
	\caption{The fidelity of the photon state (blue) and the concurrence of the system (red), representing the entanglement of the qubit and photons, are plotted as a function of the ratio $g/G$ with $G=0.1$. The inserted plots demonstrate the real part of the density matrix of the system with $g=5G$ and $35G$, respectively. Here, $|E\rangle\equiv|{\rm even}\rangle$ ($|O\rangle\equiv|{\rm odd}\rangle$) represent  the even (odd) photon number  state.}~\label{Figs:Fig7}
\end{figure}

The fidelity of the photon state can be evaluated according to the formula ${\cal F}_{\rm ph}=\langle{\rm ph}|\rho_{\rm T}|{\rm ph}\rangle$ with target density matrix of photons $\rho_{\rm T}=|{\rm even}\rangle\langle{\rm even}|+|{\rm odd}\rangle\langle{\rm odd}|$. Here, the state $|{\rm ph}\rangle$ represents the  photon state by solving the master equation. As shown in Fig.~\ref{Figs:Fig7}, the fidelity of the photon state (blue) is always close to its maximum (i.e., ${\cal F}_{\rm ph}>0.95$). To show the entanglement between qubit and cavity photons quantitatively, we can define two artificial qubits, where the first qubit has two states $|e\rangle$ and $|g\rangle$, while the second qubit has two states $|{\rm even}\rangle$ and $|{\rm odd}\rangle$. Thus, the density matrix of the system can be reconstructed by using $|g\rangle|{\rm even}\rangle$, $|g\rangle|{\rm odd}\rangle$, $|e\rangle|{\rm even}\rangle$ and $|e\rangle|{\rm odd}\rangle$ as a set of new basis. Using the reconstructed density matrix $\rho'$, the concurrence of the system (red) $C(\rho')$~\cite{note3}, characterizing the entanglement of the qubit and photons is larger than $0.93$. As the qubit-cavity coupling strength increases, both the fidelity of photon state and the concurrence between qubit and photons will be significantly improved. For weak qubit-cavity coupling strength, although the photon state can be well produced the entanglement is not very large since the energy transfer between qubit and cavity cannot be sufficiently established. To show this point, we take $g=5G$ and plot the real part of the density matrix Re$(\rho)$ in panel (c) (see inserted figures). It is clear to see that the population in state $|e,E\rangle$ is slightly larger than that in state $|g,O\rangle$. For $g=35G$, however, populations in state $|e,E\rangle$ and $|g,O\rangle$ have the same weights, corresponding to a maximum entanglement between qubit and cavity photons. 

In conclusion, we have brought out many new important consequences of the JC model when the cavity is incorporated by a parametric nonlinearity. The parametric nonlinearity produces avoided crossings between the dressed states of the JC model which otherwise cross at certain values of the detuning and the coupling of the qubit to the cavity field. At the avoided crossing one has well defined coherence between the symmetric and the antisymmetric combinations of the dressed states which leads to quantum beats in physical parameters like the excitation of the qubit. Another important feature arising from avoided crossing is the preparation of the equally weighted superpositions of the Fock states. More specifically we have shown generation of superpositions of even[ or odd] states. The fidelity of such a generation is more than 95\%. The challenging task of preparing equally weighted superpositions is achieved via the physics of the avoided crossings in the context of the cavity QED. We also demonstrate a very high degree of entanglement (concurrence more than 90\%) between the cavity field and the qubit.

\begin{acknowledgments}
CJZ and YPY thank the National Nature Science Foundation (Grant Nos. 61975154, 11774262). GSA thanks the support of Air Force Office of  scientific Research (Award No. FA9550-20-1-0366).
\end{acknowledgments}

\bibliography{Refs-QB}

\begin{thebibliography}{43}
\bibitem [{not({\natexlab{b}})}]{note3}%
  \BibitemOpen
  \bibinfo {note} {$C(\rho')={\rm
  max}(0,\lambda_1-\lambda_2-\lambda_3-\lambda_4)$. Here $\lambda_i$'s are
  square roots of eigenvalues of the matrix $\rho'\tilde{\rho'}$ in decreasing
  order. $\tilde{\rho'}$ ($\rho'^{*}$) means the transpose (complex
  conjugation) of $\rho'$.}\BibitemShut {Stop}%
\end{thebibliography}

\end{document}